\documentclass[prb, twocolumn]{revtex4-2}

\usepackage{amsmath,amsfonts,amsbsy,amsthm}
\usepackage{graphicx}
\usepackage{xcolor}

\newcommand{\sub}[1]{_{\mathrm{#1}}}
\newcommand{\su}[1]{^{\mathrm{#1}}}

\newcommand{\eps}{E}

\newcommand{\eu}{\mathrm{e}}
\newcommand{\iu}{\mathrm{i}}   %%%%% Imaginary unit
\newcommand{\di}{\mathrm{d}}
\newcommand{\BZ}{\mathrm{BZ}}

\newcommand{\up}{\uparrow}
\newcommand{\down}{\downarrow}

\newcommand{\bx}{\mathbf{R}}
\newcommand{\br}{\mathbf{R}}
\newcommand{\bk}{\mathbf{k}}

\newcommand{\bX}{\mathbf{X}}
\newcommand{\bE}{\mathbf{E}}
\newcommand{\bJ}{\mathbf{J}}
\newcommand{\bQ}{\mathbf{Q}}
\newcommand{\bK}{\mathbf{K}}
\newcommand{\bD}{\mathbf{D}}
\newcommand{\bsa}{\boldsymbol{a}}
\newcommand{\bsb}{\boldsymbol{b}}
\newcommand{\bsd}{\boldsymbol{d}}

\newcommand{\Z}{\mathbb{Z}}
\newcommand{\R}{\mathbb{R}}
\newcommand{\C}{\mathbb{C}}

\newcommand{\bra}[1]{\left\langle #1 \right|}
\newcommand{\ket}[1]{\left| #1 \right\rangle}
\newcommand{\set}[1]{ \left\{  #1 \right\}} 

\DeclareMathOperator{\Tr}{Tr}         %  Hilbert space trace

\DeclareMathOperator{\re}{Re} \DeclareMathOperator{\im}{Im}

\usepackage[normalem]{ulem}
\newcommand{\stkout}[1]{\ifmmode\text{\sout{\ensuremath{#1}}}\else\sout{#1}\fi}

\begin{document}
	
	\title{Spin Hall conductivity in insulators with non-conserved spin}
		
	\author{Domenico Monaco}
	\affiliation{Dipartimento di Matematica, La Sapienza Universit\`{a} di Roma, Piazzale Aldo Moro 5, 00185 Roma, Italy}
	\email{monaco@mat.uniroma1.it}
	
	\author{Lara Ul\v{c}akar}
	\affiliation{Department of Theoretical Physics, Jo\v{z}ef Stefan Institute, Jamova cesta 39, 1000 Ljubljana, Slovenia}
	\email{lara.ulcakar@ijs.si}
	
	\begin{abstract}
We study the linear response of a spin current to a small electric field in a two-dimensional crystalline insulator with non-conserved spin. We adopt the spin current operator proposed in [J.~Shi \textsl{et al.}, Phys.~Rev.~Lett.\ {\bf 96}, 076604 (2006)], which satisfies a continuity equation and fits the Onsager relations. We use the time-independent perturbation theory to present a formula for the spin Hall conductivity, which consists of a ``Chern-like'' term, reminiscent of the Kubo formula obtained for the quantum Hall systems, and a correction term that accounts for  the non-conservation of spin. We illustrate our findings on the Bernevig--Hughes--Zhang model and the Kane--Mele model for time-reversal symmetric topological insulators and show that the correction term scales quadratically with the amplitude of the spin-conservation-breaking terms. In both models, the spin Hall conductivity deviates from the quantized value when spin is not conserved.
\end{abstract}
	
	\maketitle
	
	\section{Introduction}
	
The spin Hall effect (SHE) has proved to be one of the essential phenomena for the manipulation of spin currents through electric fields in spintronics \cite{Schliemann06, JungwirthWunderlichOlejnik12, Sinova15}. It is present in topological insulators with time-reversal symmetry \cite{KM05, HasanKane10} that exhibit a quantized value of the spin Hall conductivity, when spin is a conserved quantity. This phenomenon was verified experimentally in HgTe nanostructures \cite{Molenkamp10} and in inverted InAs/GaSb quantum wells~\cite{InAsExp}. 

The spin transport properties of systems in which the projection of spin is not a conserved quantity have been a centre of multiple disagreements. First, recent studies~\cite{YangChang06,Ulcakar18,Bechstedt_et_al19} suggest that the quantization of the spin Hall conductivity fails when the spin is not conserved, while Ref.~\cite{Murakami06} predicts an almost quantized value. Second,
the definition of the spin current operator has been questioned, as the commonly used $\frac{1}{2} \{ \dot{\bX}, S^z\}$ does not satisfy the continuity equation in systems with non-conserved spin. Recently, a spin current operator, which evades this issue, was proposed in Ref.~\cite{Niu06} and rederived in Refs.~\cite{Tokatly08, GoriniRaimondiSchwab12} from a $SU(2)$ gauge field description of the spin degrees of freedom: this operator reads instead $\bJ^z = \iu [H_0, \bX S^z]$. 

We attempt here to resolve some of the controversies regarding the spin Hall conductivity in systems with non-conserved spin. The focus in this paper will be on the intrinsic, direct SHE, where a transverse spin current is measured as a response to an induced electric field in a two-dimensional (2D) electron gas.  In order to compute the spin Hall conductivity, we follow a time-independent perturbation scheme for linear response, inspired by space-adiabatic perturbation theory \cite{Teufel03} and its recent development in the context of charge and spin transport advanced in \cite{Teufel19, MarcelliPanatiTauber19, MarcelliMonacoPanatiTeufel}. This approach identifies the perturbed ground state projection  as a power series in the strength of the induced electric field. As in \cite{MarcelliMonacoPanatiTeufel}, we calculate the spin conductivity tensor by using the %recently proposed
spin current operator from Ref.~\cite{Niu06}, which leads to a formula for the spin Hall conductivity consisting of two contributions: the first is in a ``Chern-like'' form \cite{MarcelliPanatiTauber19} reminiscent of the standard expression for the quantum Hall conductivity, as conventionally derived from the Kubo formula \cite{TKNN82,AvronSeilerSimon83}; the second one accounts for the spin non-conserving terms in the Hamiltonian and vanishes if the spin is conserved. On the examples of the Bernevig--Hughes--Zhang (BHZ) model \cite{BHZ06} and Kane--Mele (KM) model \cite{KM05} we identify the quadratic scaling of this second contribution as a function of the spin-conservation breaking terms, and discuss how the presence of this extra contribution breaks quantization of the spin Hall conductivity in time-reversal symmetric topological insulators.

The paper is structured as follows. Sec.~\ref{sec:Theory} presents a streamlined version of the arguments from \cite{MarcelliMonacoPanatiTeufel}: in particular, in Sec.~\ref{sec:EqSt} we characterize the studied systems, introduce the spin current in Sec.~\ref{sec:SpinCur} and the perturbed state in the framework of time-independent perturbation theory in Sec.~\ref{sec:Perturbed}; finally we derive the formula for the spin Hall conductivity in Sec.~\ref{sec:SHC}. In Sec.~\ref{sec:Numerics} we explain the numerical evaluation of the formula, and present the results for the BHZ model and the KM model in Sec.~\ref{sec:models}.

\section{Linear response for spin currents} \label{sec:Theory}

\subsection{Equilibrium state} \label{sec:EqSt}
We study a tight-binding Hamiltonian $H_0$ acting on states $\ket{\bx \, s \, \sigma}$, where $\bx$ is a vector in a 2D Bravais lattice, labelling cells in a crystal consisting of $N_1 \times N_2$ copies of the fundamental cell $\Omega$ with odd $N_1, N_2$, $s \in \set{\up,\down}$ is the physical spin-$1/2$ and $\sigma$ includes all other local degrees of freedom in the unit cell. We denote in particular by $\bX \ket{\bx \, s \, \sigma} = (\bx+\mathbf{r}_\sigma) \ket{\bx \, s \, \sigma}$ the vector of position operators ($\mathbf{r}_\sigma$ is a displacement vector inside the unit cell) and by $S^a = s^a / 2$, for $a \in \set{x,y,z}$, the spin operators acting as half of the Pauli matrices on the spin degree of freedom. Throughout the paper we adopt units in which $\hbar = 1$. The Hamiltonian $H_0$ is assumed to be translation invariant and gapped: the Fermi energy lies in the spectral gap, and $\Pi_0$ denotes the ground state Fermi projection onto occupied states. In the numerical results described in Section~\ref{sec:models}, $H_0$ will be specified to be the BHZ and KM Hamiltonian at half-filling. We stress that $H_0$ may contain terms, which do not conserve the spin (or rather its projection along the $z$-direction), like Rashba spin-orbit coupling: $[H_0,S^z] \ne 0$.

Since the system is translation invariant, its Hamiltonian and the ground-state projector $\Pi_0$ are of  diagonal form when written in the Fourier-transformed basis $\ket{\bk \, s \, \sigma} = (1/ \sqrt{N_1 \, N_2}) \sum_{\bx} \eu^{\iu \bk \cdot (\bx+\mathbf{r}_\sigma)} \ket{\bx \, s \, \sigma}$% {\red [CHECK that this is the correct definition -- see also \eqref{eqn:partialFourier}. It is \textbf{not} periodic w.r.t.\ the dual lattice: $\ket{\bk+\mathbf{G} \, s \, \sigma} = \eu^{\iu \mathbf{G} \cdot \mathbf{r}_\sigma} \ket{\bk+\mathbf{G} \, s \, \sigma}$ -- is this a problem?]} \textcolor{blue}{No, without $e^{i\mathbf{k\cdot r}_\sigma}$}
. For example, $\Pi_0$ in the Fourier representation is of the form
\begin{equation} \label{eqn:Pi0}
\begin{gathered}
\Pi_0 = \sum_\bk \ket{\bk \, s' \, \sigma'} \Pi_0(\bk)_{s\,\sigma}^{s'\,\sigma'} \bra{\bk \, s \, \sigma},\\
\Pi_0(\bk)=\sum_{\alpha \text{ occupied}} \ket{\psi_{\alpha \bk}} \bra{\psi_{\alpha \bk}},
\end{gathered}
\end{equation}
where $\ket{\psi_{\alpha \bk}} \in \C^n$ is the Bloch function associated to the $\alpha$-th occupied band at crystal momentum $\bk$, and $n$ is the number of degrees of freedom per cell. $H_0(\bk)$ and the corresponding $\Pi_0(\bk)$  act as $n\times n$ matrices on spin and orbital degrees of freedom.% and they are periodic with respect to momentum shifts in the reciprocal lattice.% {\red [With the above definition of Fourier transform, $\Pi_0(\bk+\mathbf{G})_{s\,\sigma}^{s'\,\sigma'} = \eu^{-\iu \mathbf{G} \cdot (\mathbf{r}_\sigma - \mathbf{r}_{\sigma'})}\Pi_0(\bk)_{s\,\sigma}^{s'\,\sigma'}$, isn't it?]} \textcolor{blue}{I believe not}%: therefore the values of $\bk$ may be restricted to the Brillouin zone (BZ).

\subsection{Spin current operator} \label{sec:SpinCur}

The SHE is observed by inducing a weak electric field and measuring the response of a spin current in the limit of small perturbation. We therefore set
\begin{equation} H_\eps = H_0 - \bE \cdot \bX 
\end{equation}
for the perturbed Hamiltonian, where the charge of the carriers is assumed to be $q=1$: the label $\eps = |\bE|$ is the strength of the inducing field, which is assumed to be small. To measure the spin response, we adopt the following definition of the spin current operator~\cite{Niu06}: $\bJ^z = \iu [H_0, \bQ]$, where $\bQ = \bX S^z$. This current operator is \emph{not} translation invariant unless spin is conserved, since
\begin{equation} \label{eq:Scurrent}
\bJ^z = \iu \bX \underbrace{[H_0,S^z]}_{\text{transl.-inv.}} + \underbrace{\iu [H_0, \bX] S^z}_{\text{transl.-inv.}}. 
\end{equation}
Nonetheless, the operator $\bJ^z$ has a well-defined expectation $\langle \bJ^z \rangle_{\Pi}$ in any translation-invariant state $\Pi$ \cite{MarcelliMonacoPanatiTeufel}, as we detail in Appendix~\ref{app:TPUV}, which can be expressed as a trace per unit volume (TPUV): $\langle \bJ^z \rangle_{\Pi} = \re \tau \left\{ \bJ^z \Pi \right\}$ with
\begin{equation} \label{eqn:Jzrho}
 \tau \left\{ A \right\} = \frac{1}{|\Omega|} %\sum_{\bx \in \Omega} 
 \sum_{s, \sigma} \bra{\mathbf{0} \, s \, \sigma} A \ket{\mathbf{0} \, s \, \sigma},
\end{equation}
where the %first sum runs
expectation values of $A$ are taken only on states localized over sites in the fundamental cell $\Omega$.
%\textcolor{red}{What is $\rho$? Is this the density matrix of the system or the projector on the state?}
	
\subsection{Perturbed state}\label{sec:Perturbed}
	
The induced electric field changes the state of the system from the equilibrium Fermi projection $\Pi_0$ to a state~$\rho_\eps$. %Treating the perturbation as space-adiabatically small, in the sense of~\cite{Teufel03}, 
	We adopt the time-independent Rayleigh--Schr\"{o}dinger perturbation scheme \cite{sakurai_napolitano_2017, MarcelliMonacoPanatiTeufel} and approximate the state~$\rho_\eps$ to the first order in the strength $\eps$ of the electric field as $\rho_\eps = \Pi_0 + \eps \, \Pi_1 + O(\eps^2)$. % characterized by the properties of being a translation-invariant orthogonal projection which almost-commutes with $H_\eps$, in the sense that $[H_\eps, \Pi_\eps] = O(\eps^2)$. 
	The first order term $\Pi_1$ is a translation-invariant self-adjoint operator characterized by the fact that it is \emph{off-diagonal} with respect to the orthogonal decomposition induced by $\Pi_0$, i.e. 
	\begin{equation} \label{eqn:Pi1OD}
	\Pi_1 = \Pi_1\su{OD} = \Pi_0 \, \Pi_1 \, (\mathbf{1} - \Pi_0) +  (\mathbf{1} - \Pi_0) \, \Pi_1 \, \Pi_0,
	\end{equation}
and that it satisfies the commutation relation
	\begin{equation} \label{eqn:CommPi1}
	[H_0, \Pi_1] = [\hat{\bE} \cdot \bX, \Pi_0] \quad \text{with} \quad \hat{\bE} = \bE / \eps.
	\end{equation}
	
By using that $[\bX, \Pi_0](\bk) = \iu \, \nabla_{\bk} \Pi_0(\bk)$, one can derive the following expression for the matrix elements of $\Pi_1(\bk)$ in the basis of Bloch states:
\begin{equation} \bra{\psi_{\alpha \bk}}\Pi_1(\bk) \ket{\psi_{\beta \bk}} = \frac{\iu \left\langle \psi_{\alpha \bk} \,\Big|\, \hat{\bE} \cdot \nabla_{\bk} \Big| \psi_{\beta \bk} \right\rangle}{\epsilon_{\beta \bk} - \epsilon_{\alpha \bk}},
\end{equation}
where $\ket{\psi_{\alpha \bk}}$ is an occupied Bloch state, $\ket{\psi_{\beta \bk}}$ is an unoccupied Bloch state, while $\epsilon_{\alpha \bk}$ and $\epsilon_{\beta \bk}$ are their respective Bloch energies. The other matrix elements of $\Pi_1(\bk)$ can be inferred by using the fact that it is self-adjoint and that it is off-diagonal. Notice that the numerator of the above expression is, up to the sign, the non-Abelian Berry connection $\mathcal{A}(\bk)_{\alpha \beta} = \im \left\langle \psi_{\alpha \bk} \,\big|\, \nabla_{\bk}\big| \psi_{\beta \bk} \right\rangle$ dotted with the direction $\hat{\bE}$ of the  electric field. The denominator is the hallmark of the inverse Liouvillian $\mathcal{L}_{H_0(\bk)}^{-1}$, where $\mathcal{L}_{H_0(\bk)}(A(\bk)) = [H_0(\bk), A(\bk)]$.

For the numerical evaluation of $\Pi_1(\mathbf{k})$ it is more suitable to use a gauge-invariant equation, which for a 4-band Hamiltonian and electric field in $y$-direction reads explicitly
\begin{equation} \label{eq:Pi1}
\begin{aligned}
&\big( \epsilon_{1\,\bk} \ket{\psi_{1\,\bk}} \bra{\psi_{1\,\bk}} + \epsilon_{2\,\bk} \ket{\psi_{2\,\bk}} \bra{\psi_{2\,\bk}} \big) \Pi_1(\bk)\sub{u-o} \\
& - \Pi_1(\bk)\sub{u-o} \big( \epsilon_{3\,\bk} \ket{\psi_{3\,\bk}} \bra{\psi_{3\,\bk}} + \epsilon_{4\,\bk} \ket{\psi_{4\,\bk}} \bra{\psi_{4\,\bk}} \big) \\
& = \iu \, [\partial_{k_y}\Pi_0(\bk)]\sub{u-o}.
\end{aligned}
\end{equation}
Here, $\Pi_1(\bk)\sub{u-o} = \Pi_0(\bk) \Pi_1(\bk) (\mathbf{1} - \Pi_0(\bk))$ is the unoccupied-to-occupied block. The equation is obtained by recombining the defining equations \eqref{eqn:Pi1OD} and \eqref{eqn:CommPi1}.
	
\subsection{Spin Hall conductivity}\label{sec:SHC}

The spin conductivity tensor $\boldsymbol{\sigma}^z$ is defined through Ohm's law:
\begin{equation}
\langle\bJ^z\rangle_{\rho_\eps} = \langle\bJ^z\rangle_{\Pi_0} +  \boldsymbol{\sigma}^z \, \bE. 
\end{equation}
In the limit of a small inducing field, only $\Pi_1$ contributes to the linear response of the spin current. The spin conductivity tensor $\boldsymbol{\sigma}^z$ can thus be evaluated via
\begin{equation}\label{eq:sHall}
 \re \tau \{ \bJ^z \, \Pi_1 \} = \boldsymbol{\sigma}^z \, \hat{\bE}. 
\end{equation}

Using the defining relation from Eq.~\eqref{eqn:CommPi1} of $\Pi_1$ and observing that $A\su{OD} = \big[ [A, \Pi_0], \Pi_0 \big]$ for any operator $A$, the expression $\bJ^z \, \Pi_1$ can be rewritten with straightforward algebraic manipulations (see Appendix \ref{app:internal}) as $\bJ^z \, \Pi_1 = \bK + \bD$, where
\begin{equation} \label{K&D}
\begin{split}
\bK & = \iu \big[ [\bQ,\Pi_0], [\hat{\bE} \cdot \bX, \Pi_0] \big] \Pi_0, \\
\bD & = \iu \left[H_0,\bQ\su{D}\right] \Pi_1 + \iu \left[H_0,\bQ\su{OD} \, \Pi_1\right] \\
& \quad + \iu \big[ [\bQ,\Pi_0],\Pi_0 [\Pi_0,\hat{\bE} \cdot \bX] \big].
\end{split}
\end{equation}
In the above formula, $\bQ\su{OD}$ and $\bQ\su{D} = \bQ - \bQ\su{OD}$ denote, respectively, the off-diagonal and the diagonal part of the operator $\bQ$ in the orthogonal decomposition induced by~$\Pi_0$ (compare~Eq.~\eqref{eqn:Pi1OD}). Correspondingly, we split the conductivity tensor as $\boldsymbol{\sigma} = \boldsymbol{\sigma}\su{I} + \boldsymbol{\sigma}\su{II}$, with the two contributions determined by the conditions
\begin{equation} 
\tau \{ \bK \} = \boldsymbol{\sigma}\su{I} \, \hat{\bE}, \quad \re \tau \{ \bD \} = \boldsymbol{\sigma}\su{II} \, \hat{\bE}.
\end{equation}

Since we are interested in the SHE, we apply the inducing electric field in the $y$-direction and consider the response of the orthogonal $x$-component of the spin current. Writing $\bX = (X,Y)$ and $\bQ = (Q_x, Q_y) = (XS^z, YS^z)$, the spin Hall conductivity $\sigma_{xy}^{z}$ evaluates then to $\sigma_{xy}^z = \sigma_{xy}\su{I} + \sigma_{xy}\su{II}$ with
\begin{equation} \label{eqn:I&II}
\begin{aligned}
\sigma_{xy}\su{I} & = \tau\! \left\{ \iu \, \big[ [Q_x,\Pi_0], [Y,\Pi_0] \big]\Pi_0 \right\}, \\
\sigma_{xy}\su{II} & = \re \: \tau \Big\{ \iu \, [H_0,Q_x\su{D}] \Pi_1 + \iu [H_0,Q_x\su{OD}\Pi_1] \\
& \qquad\qquad + \iu \big[ [Q_x,\Pi_0],\Pi_0 [\Pi_0,Y] \big] \Big\}.
\end{aligned}
\end{equation}

The first term $\sigma_{xy}\su{I}$ in Eq.~\eqref{eqn:I&II} is a ``Chern-like'' contribution to the spin Hall conductivity, while $\sigma_{ij}\su{II}$ accounts for extra contributions coming from the spin non-conserving terms in the Hamiltonian $H_0$, and vanishes if the spin is conserved. Indeed, if $[H_0,S^z] = 0$, one has that $Q_x\su{D/OD} = X\su{D/OD} S^z$, because $S^z$ commutes with $\Pi_0$. Then the operator $\iu \left[H_0,Q_x\su{D}\right] \Pi_1 = \iu [H_0,X]\su{D} S^z \Pi_1$ is translation invariant and is a product of the diagonal operator $\iu [H_0,X]\su{D} S^z$ with the off-diagonal one $\Pi_1$: hence it is off-diagonal and has no diagonal matrix elements, so that its TPUV computed as in Eq.~\eqref{eqn:Jzrho} vanishes. Similarly, the other two contributions to $\bD$, namely
\begin{equation}
\iu \left[H_0,Q_x\su{OD}\Pi_1\right] = \iu \left[H_0,X\su{OD} S^z\Pi_1\right]
\end{equation}
and
\begin{equation}
\iu \big[ [Q_x,\Pi_0],\Pi_0 [\Pi_0,Y] \big] = \iu \big[ [X,\Pi_0] S^z,\Pi_0 [\Pi_0,Y] \big],
\end{equation}
are translation invariant and in the form of commutators, so they also have vanishing TPUV. We conclude that, if the spin is conserved, then $\tau \{ \bD \} = 0$, and the expression for the spin Hall conductivity, consisting only of the $\sigma_{xy}\su{I}$ term that involves translation invariant operators, can be written in momentum space and in the thermodynamic limit as
\begin{align}
\sigma_{xy}^z & = \frac{-\iu}{(2 \pi)^2} \intop_{\BZ} \di \bk \Tr_{\C^n} \left( S^z \Pi_0(\bk) \big[ \partial_{k_x} \Pi_0(\bk), \partial_{k_y} \Pi_0(\bk) \big] \right) \nonumber \\
& = \frac{1}{2}\left(c_1(\Pi_0^\up)-c_1(\Pi_0^\down)\right),
\end{align}
where $\Pi_0^{\up/\down}$ are the restrictions of the Fermi projection to the spin-up/spin-down sectors, respectively, and $c_1(\Pi)$ is the first Chern number associated to a family of projections $\Pi(\bk)$ \cite{AvronSeilerSimon83,AvronSeilerSimon94}. The above coincides with the known formula for the spin Hall conductivity \cite{Kane05} and gives the spin Chern number~\cite{Prodan09} that is half-quantized in units of $\frac{1}{2\pi}$. Under the further assumption of time-reversal symmetrys on $H_0$, which implies in particular $c_1(\Pi_0^\down) = - c_1(\Pi_0^\up)$, the parity of the integer $2 \pi \sigma_{xy}^z$ coincides with the Kane--Mele $\Z_2$ index~\cite{KM05,FuKane06}.

When instead $[H_0, S^z] \ne 0$, the extra terms in $\sigma_{xy}^z$ coming from $\bD$ are still off-diagonal or in the form of commutators, however they are not translation invariant. It turns out that, for non-translation-invariant operators, the TPUV may fail to be \emph{cyclic}, that is, it is not true in general that $\tau(AB) = \tau(BA)$. This means that $\sigma_{xy}\su{II}$ may be non-zero, and we will see this to be the case in both the BHZ and KM models when the spin is not conserved.
	
	\section{Numerical implementation}\label{sec:Numerics}

Let $\boldsymbol{a}_1$ and $\boldsymbol{a}_2$ be the primitive vectors of the Bravais lattice with which we can describe the position of an arbitrary unit cell as $\br=R_1\boldsymbol{a}_1+R_2\boldsymbol{a}_2$ with integer $R_1, R_2$. On the finite lattice, each $R_j$ ranges from $-(N_j-1)/2$ to $(N_j-1)/2$, for $j \in \set{1,2}$. The reciprocal lattice is spanned by reciprocal vectors $\boldsymbol{b}_1$ and $\boldsymbol{b}_2$ defined by $\boldsymbol{a}_i\cdot \boldsymbol{b}_j=2\pi\delta_{ij}$. The momentum $\bk=k_1\boldsymbol{b}_1+k_2\boldsymbol{b}_2$ is an element of the discrete 2D Brillouin zone, with coordinates $k_j\in\mathrm{BZ_j}$ where $\BZ_j = \set{-\frac{N_j-1}{2N_j}, \ldots,\frac{N_j-1}{2N_j}}$ for $j=1,2$.
	
In order to compute the spin Hall conductivity $\sigma_{xy}^z$, we turn on the electric field in the $y$-direction, which will also be aligned with the vector $\boldsymbol{a}_2$. As seen from Eq.~\eqref{eqn:I&II}, the expression for $\sigma_{xy}^z$ involves operators which are not translation invariant in the $x$-direction. Therefore we introduce the partial Fourier transform relating the $\ket{\bk \, s \, \sigma}$-basis to the $\ket{(R_1,k_2) \, s \, \sigma}$-basis as follows:
\begin{equation}\label{eqn:partialFourier}
\ket{\bk \,s \, \sigma} = \frac{1}{\sqrt{N_1}} \sum_{R_1} c_{\sigma}(k_1)\,\eu^{\iu 2\pi k_1 R_1} \ket{(R_1,k_2) \, s \, \sigma}.
\end{equation}
The coefficient $c_\sigma(k_1) = \eu^{\iu k_1 \bsb_1 \cdot \mathbf{r}_\sigma}$ accounts for the displacement $\mathbf{r}_\sigma$ of the local degrees of freedom in the unit cell: we will specify it later in the concrete models. 
Conveniently, the position operator $X$ is diagonal in this basis: 
\begin{equation} \label{eqn:Xsigma}
X=\sum_{R_1,k_2,s,\sigma}x_{\sigma}(R_1)\ket{(R_1,k_2)\,s\,\sigma}\bra{(R_1,k_2)\,s\,\sigma},
\end{equation}
where the value of $x_\sigma(R_1) = (R_1 \bsa_1 + \mathbf{r}_\sigma) \cdot \boldsymbol{\hat{e}}_x$ will again be specified later in the BHZ and KM models.
	
To calculate $\sigma_{xy}^z$, we express all operators (e.g.~$H_0(\bk)$, $\Pi_0(\bk)$, $\Pi_1(\bk)$) as matrices in the new basis (correspondingly $H_0(k_2)$, $\Pi_0(k_2)$, $\Pi_1(k_2)$): they act now also on the position $x$ degree of freedom. The formula for the two contributions to the spin Hall conductivity appearing in Eq.~\eqref{eqn:I&II} is expressed in the $\ket{(R_1,k_2) \, s \, \sigma}$-basis as
\begin{widetext}
	\begin{equation} \label{eqn:I&IIxkKM}
	\begin{aligned}
		\sigma_{xy}\su{I} & = - \frac{1}{|\Omega| \, N_2} \sum_{k_{2} \in \BZ_2} \sum_{s \in \set{\up,\down}} \sum_{\sigma} \bra{(0,k_2) \, s \, \sigma} \big[ [Q_x,\Pi_0(k_2)], \partial_{k_y}\Pi_0(k_2) \big] \Pi_0(k_2) \ket{(0,k_2) \, s \, \sigma} \, ,\\
		\sigma_{xy}\su{II} & = -\frac{1}{|\Omega| \, N_2} \sum_{k_2 \in \BZ_2} \sum_{s \in \set{\up,\down}} \sum_{\sigma} \im \bra{(0,k_2) \, s \, \sigma} \Big\{ \left [H_0(k_2),Q_x\su{D} \right] \Pi_1(k_2) + \left[H_0(k_2),Q_x\su{OD}\Pi_1(k_2)\right] \\
		& \quad + \big[ [Q_x,\Pi_0(k_2)],\Pi_0(k_2) (-\iu \partial_{k_y}\Pi_0(k_2)) \big] \Big\} \ket{(0,k_2) \, s \, \sigma} \, ,
\end{aligned}
\end{equation}
\end{widetext}
Notice that in the above expressions the derivative with respect to $k_y$ appears, due to the fact that the electric field is applied in the $y$-direction.

\section{Results for reference models}
\label{sec:models}

\subsection{Bernevig-Hughes-Zhang model} \label{sec:BHZmodel}

\subsubsection{The model}

The BHZ Hamiltonian was introduced in~\cite{BHZ06} to model the low-energy physics of HgTe/CdTe quantum wells that realize time-reversal symmetric topological insulators, in which the quantum spin Hall effect can take place~\cite{Molenkamp10}. In systems with band inversion asymmetry and structural inversion asymmetry, such as InAs/GaSb/AlSb Type-II semiconductor quantum wells, terms that couple states with opposite spin projections and preserve the time-reversal symmetry arise. The tight-binding model takes place on a square crystalline lattice with two orbitals per lattice site. We set the lattice constant to $1$. The $\bk$-space expression for the BHZ Hamiltonian reads
\begin{align}
H_0(\bk) & = s^0 \otimes \big[ (u + \cos(k_x) + \cos(k_y)) \,\sigma^z + \sin(k_y) \, \sigma^y \big]\nonumber \\
& \quad + s^z \otimes \sin(k_x) \, \sigma^x + c \, s^x \otimes \sigma^y,
\end{align}
where $(k_x,k_y) = (2\pi k_1,2\pi k_2)$ are the Cartesian coordinates of momentum, $s^a$ and $\sigma^a$ for $a \in \{x,y,z\}$ are Pauli matrices in spin space and local orbital space, respectively, and $s^0$ is the identity operator in spin space.
The Hamiltonian above is expressed in units of the inter-cell hopping amplitude, which is equal in both directions. The parameter $u \in \R$ is the staggered orbital binding energy, while $c \in \R$ is the coupling constant between spin and orbital degrees of freedom: the latter determines the magnitude of the spin conservation breaking as $[H_0(\bk),S^z] \propto c$. When $c=0$, the original BHZ model is recovered. The topological phase diagram, characterized by the $\mathbb{Z}_2$ index, is discussed in Ref.~\cite{Ulcakar18}. In the topological phase the system exhibits the SHE.

\subsubsection{Spin Hall conductivity}

\begin{figure*}%[h!]
	\centerline{
		\includegraphics[width=130pt]{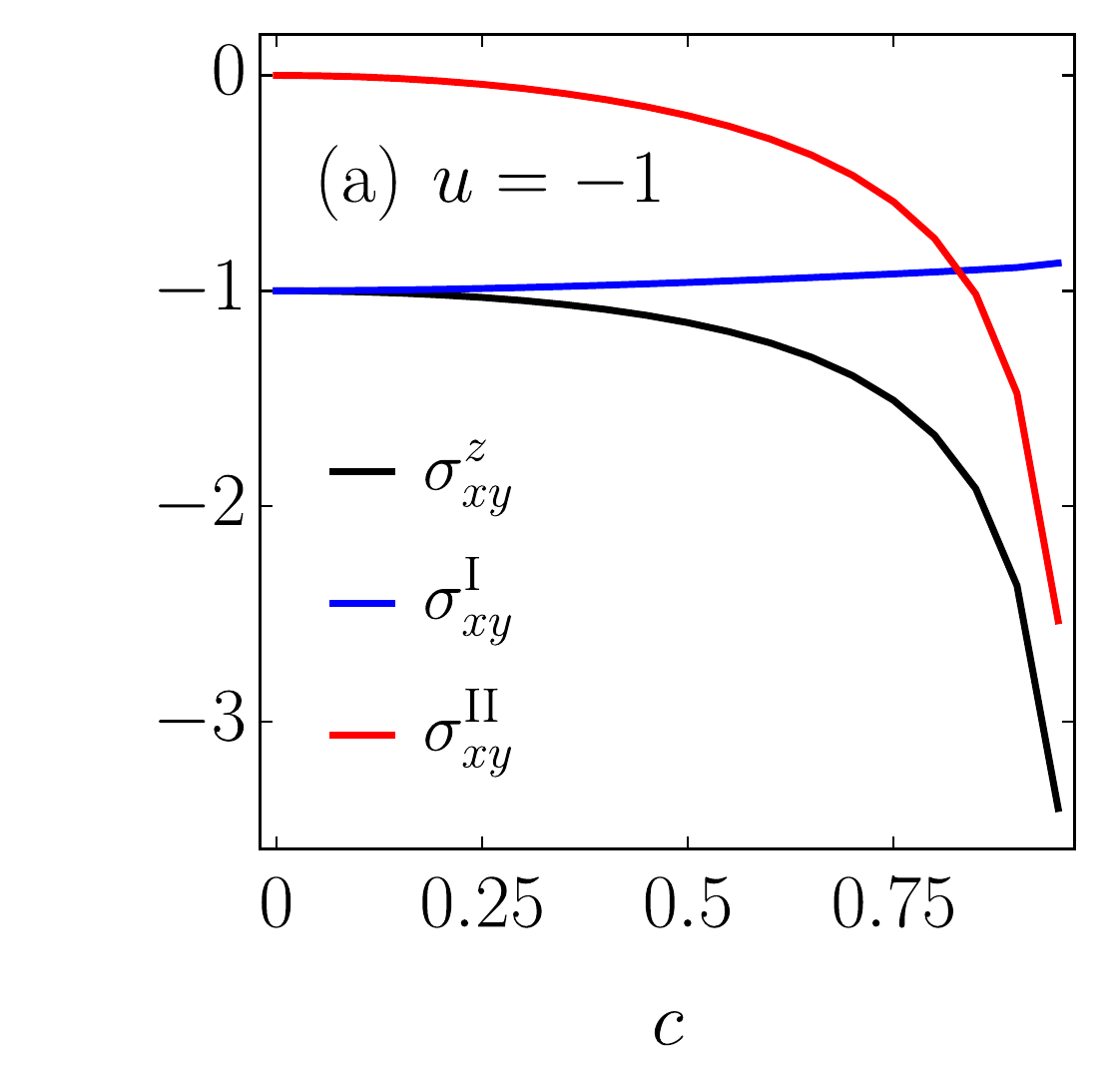}\includegraphics[width=130pt]{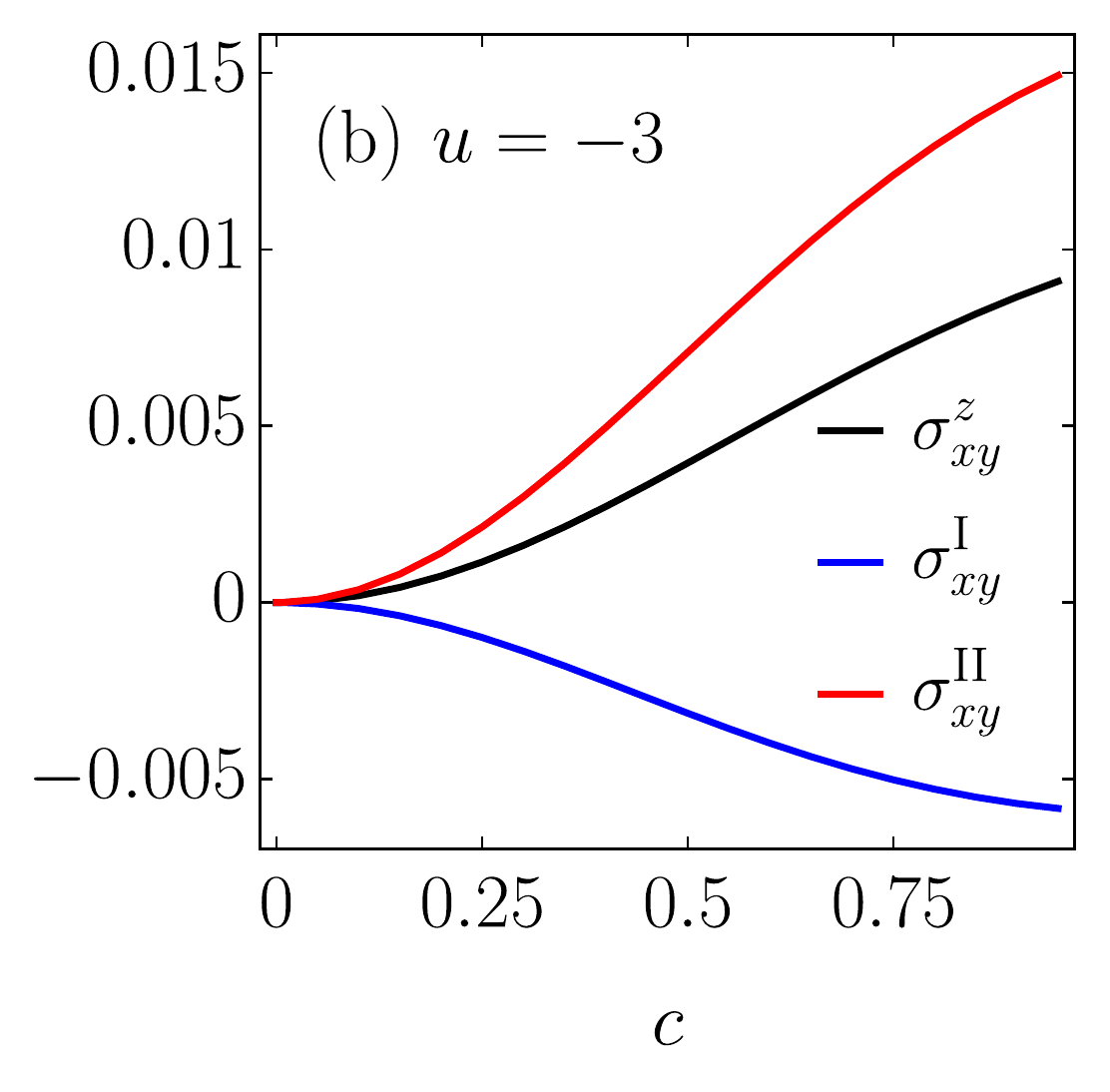}\includegraphics[width=130pt]{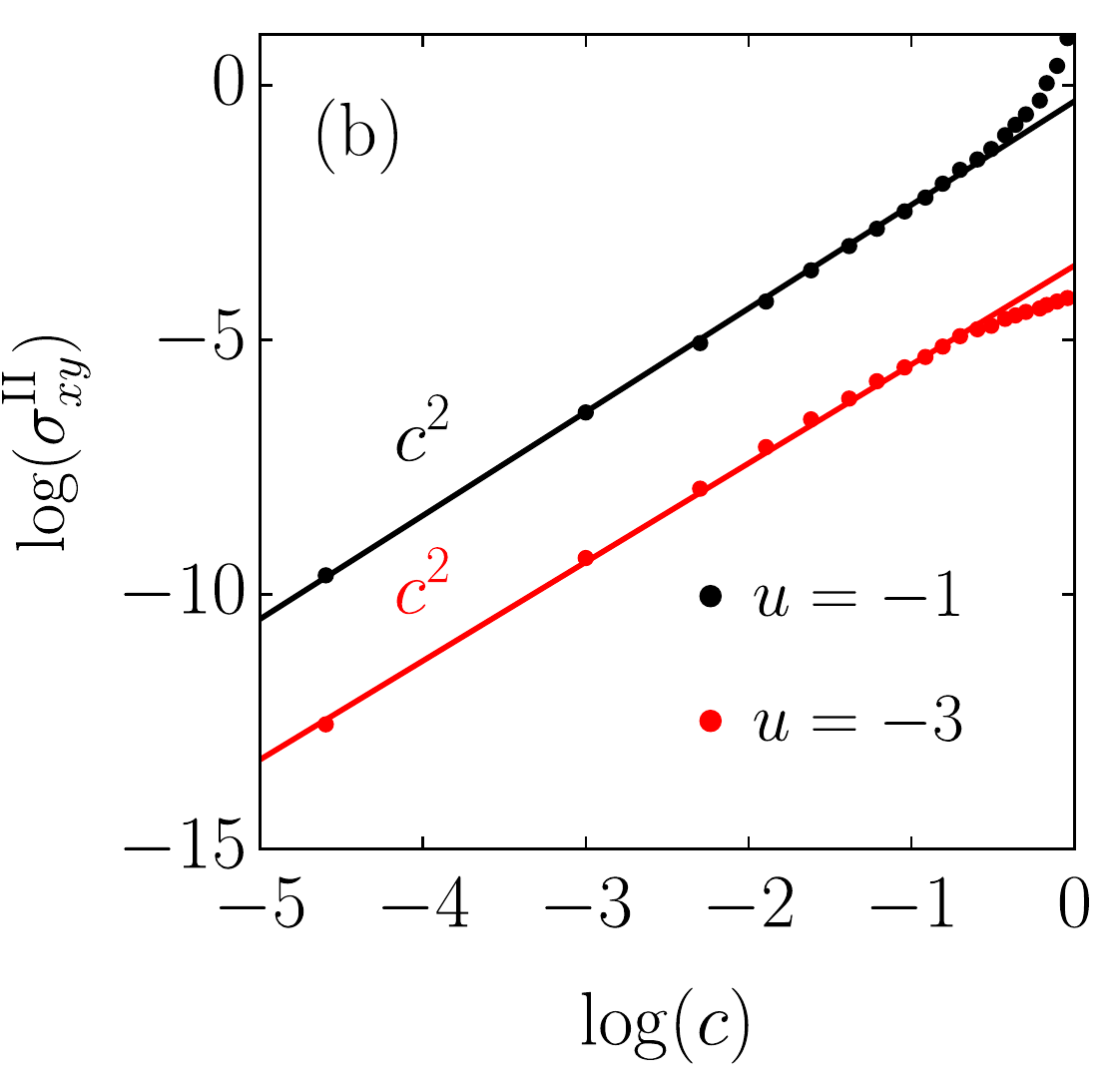}
	}
	\caption{Spin Hall conductivity $\sigma_{xy}^{z}$, the ``Chern-like'' contribution $\sigma_{xy}\su{I}$ and the ``extra'' contribution $\sigma_{xy}\su{II}$ in the BHZ model, computed on a $31 \times 31$ finite lattice. 
		Dependence on $c$ for the system in (a) the topological phase at $u=-1$ and in (b) the trivial phase at $u=-3$ and (c) the scaling of $\sigma_{xy}\su{II}$ with $c$ on a log-log scale. Dots show numerical results while full lines denote fitted $c^2$ curve.}
	\label{fig:SxyOfC}
\end{figure*}

To compute the spin Hall conductivity as in Eq.~\eqref{eqn:I&IIxkKM}, we need to set $c_\sigma(k_1)=1$ in Eq.~\eqref{eqn:partialFourier} and $x_\sigma(R_1)=R_1$ in Eq.~\eqref{eqn:Xsigma}, since all of the internal degrees of freedom are placed on the same position in a unit cell. 
The numerical results for $\sigma_{xy}\su{I}$, $\sigma_{xy}\su{II}$ and $\sigma_{xy}^{z}$ are plotted in Figure~\ref{fig:SxyOfC}. Both $\sigma_{xy}\su{I}$ and $\sigma_{xy}\su{II}$ show a non-trivial dependence on the spin coupling $c$, as illustrated by Fig.~\ref{fig:SxyOfC}(a)--(b). As expected, at $c=0$ the spin Hall conductivity coincides with the ``Chern-like'' contribution $\sigma_{xy}\su{I}$ and equals $0$ in the trivial phase and $-1$ in the topological phase. For $c\neq 0$, the spin Hall conductivity deviates from the quantized value: the change is mostly due to the increase in magnitude of $\sigma_{xy}\su{II}$, which scales quadratically in $c$ (Fig.~\ref{fig:SxyOfC}(c)), while $\sigma_{xy}\su{I}$ stays almost unchanged in relative size.

Interestingly, our results coincide with the results of Ref.~\cite{Ulcakar18}, which uses the conventional definition for the spin current $\bJ^z\sub{conv} = \frac{1}{2} \{ \dot{\bX}, S^z\}$. 
As proved in Appendix~\ref{sec:conv=prop}, this is the consequence of the lattice structure, namely because the orbital degrees of freedom are all positioned on the same lattice site (see also \cite{MarcelliMonacoPanatiTeufel}).

\subsection{Kane-Mele model} \label{sec:KMmodel}

\subsubsection{The model}

The KM Hamiltonian was introduced in~\cite{KM05} as a candidate model for the quantum spin Hall effect in graphene. Electrons reside on the hexagonal lattice with hopping parameter $t$ and are subject to spin-orbit interaction with coupling strength $\lambda\sub{SO}$, Rashba interaction with coupling $\lambda\sub{R}$ and to a staggered potential, equal to $\pm M$ on neighbouring sites. 

The honeycomb lattice is made of two interpenetrating Bravais triangular sublattices, commonly denoted by $A$ and $B$. We set the lattice constant of a Bravais lattice to $1$. Starting from an $A$-site %(dark in Figure~\ref{fig:latticeKM}) 
as the origin, the nearest-neighbour (NN) sites are of $B$-type and they are reached with the three displacement vectors
\begin{equation}
\bsd_1 = \frac{1}{2}\left( \frac{1}{\sqrt{3}}, -1 \right), \bsd_2 = \frac{1}{2}\left( \frac{1}{\sqrt{3}},1 \right),\bsd_3 = -\bsd_1-\bsd_2,
\end{equation}
where $\left|\bsd_i\right|=1/\sqrt{3}$ is the NN distance. We denote the primitive vectors of the Bravais lattice as
\begin{equation}
\bsa_1 = \left( \frac{\sqrt{3}}{2} , \frac{1}{2} \right), \quad \bsa_2 =  (0,1),\quad \bsa_3 = \bsa_2 - \bsa_1. 
\end{equation}
The unit cell is generated by $\bsa_1$ and $\bsa_2$, and contains one $A$-site and one NN $B$-site, which constitute the internal degree of freedom denoted by $\sigma\in\{A,B\}$. %\textcolor{red}{(Would not include the following paragraph in the submitted version.)} 

The reciprocal lattice vectors $\bsb_1$ and $\bsb_2$ are constructed in the standard way by imposing $\bsa_i \cdot \bsb_j = 2 \pi \delta_{ij}$:
\begin{equation}
\boldsymbol{b}_1 = \frac{4\pi}{\sqrt{3}} \, (1,0), \quad \boldsymbol{b}_2 = \frac{4\pi}{\sqrt{3}} \, \left(-\frac{1}{2}, \frac{\sqrt{3}}{2}\right).
\end{equation}
%We define two-component spinors $\ket{\bk \, \sigma} = \big( \ket{\bk \, \up \, \sigma}, \ket{\bk \, \down \, \sigma} \big)$, $\sigma \in \set{A,B}$, as \textcolor{red}{why do we need these spinors?}% \textcolor{red}{is this corrent? k points are defined without the $2\pi$ factor so there should be one in the exponent in Bloch waves below?}
%\begin{align*}
%\ket{\bk \, A} & = \frac{1}{\sqrt{N_1 N_2}} \sum_{\bx \in A} \eu^{\iu \bk \cdot \bx} \ket{\bx} = \frac{1}{\sqrt{N_1 N_2}} \sum_{\br \in A} \eu^{\iu \bk \cdot \br} \ket{\br \, A}, \\
%\ket{\bk \, B} & = \frac{1}{\sqrt{N_1 N_2}} \sum_{\bx \in B} \eu^{\iu \bk \cdot \bx} \ket{\bx} \\
%& = \frac{1}{\sqrt{N_1 N_2}} \sum_{\br \in A} \eu^{\iu \bk \cdot (\br + \bsd_2)} \ket{\br \, B},
%\end{align*}
%where
%\begin{align*}
%\ket{\br \, A} & = \big( \ket{\br \, \up \, A}, \: \ket{\br \, \down \, A} \big), \\
%\ket{\br \, B} & = \big( \ket{\br+\bsd_2 \, \up \, B}, \: \ket{\br+\bsd_2 \, \down \, B} \big).
%\end{align*}
When periodic boundary conditions are imposed, the system is translation invariant and the Hamiltonian can be expressed in momentum space as
\begin{equation}
\begin{split}
H_0(\bk) & =
\frac{1}{2} \big(M s^0 + \lambda\sub{SO} \gamma(\bk)  s^z \big)\otimes\sigma^z \\
& +\big[ t g(\bk) s^0 + \lambda\sub{R} \big( \chi_x(\bk) s^x - \iu\chi_y(\bk) s^y \big) \big] \\
&\otimes\frac{\sigma^x+\iu\sigma^y}{2} +\mathrm{h.c.}
\end{split}
\end{equation}
with
\begin{equation}
\begin{split}
&g(\bk)  = - \sum_{i=1}^{3} \eu^{\iu \bk \cdot \bsd_i}, \\
&\gamma(\bk)  = - 2 \,\sum_{i=1}^{3} \sin(\bk \cdot \bsa_i), \\
&\chi_x(\bk)  = \frac{\iu \,\sqrt{3}}{2} (  \eu^{\iu \bk \cdot \bsd_1}  - \eu^{\iu \bk \cdot \bsd_2}), \\
&\chi_y(\bk)  = -\frac{1}{2}( \eu^{\iu \bk \cdot \bsd_1} + \eu^{\iu \bk \cdot \bsd_2} - 2  \, \eu^{\iu \bk \cdot \bsd_3} ).
\end{split}
\end{equation}

%\textbf{[The discussion of these symmetries is needed ONLY if we keep the discussion of proper vs conventional spin currents]} Time-reversal symmetry is expressed by the fact that 
%\[ T \, H_0\su{KM}(\bk)^* \, T^{-1} = H_0\su{KM}(-\bk), \]
%where $T = \eu^{\iu \pi S^y}$. From the above expression, it is also seen that the mirror symmetry $(x,y) \mapsto (x,-y)$ of the hexagonal lattice translates into the relation
%\[ T \, H_0\su{KM}(k_x,k_y) \, T^{-1} = H_0\su{KM}(k_x,-k_y). \]
%These two relations can be combined into the symmetry
%\begin{equation} \label{hidden}
%H_0\su{KM}(k_x,k_y)^* = H_0\su{KM}(-k_x,k_y),
%\end{equation} 
%which reflects the fact that a mirror symmetry $(x,y) \mapsto (-x,y)$ of the lattice transforms it into a new hexagonal lattice, shifted by $\bsd_2$ with respect to the original one.

\subsubsection{Spin Hall conductivity}
\begin{figure*}%[h!]
	\centerline{\includegraphics[width=260pt]{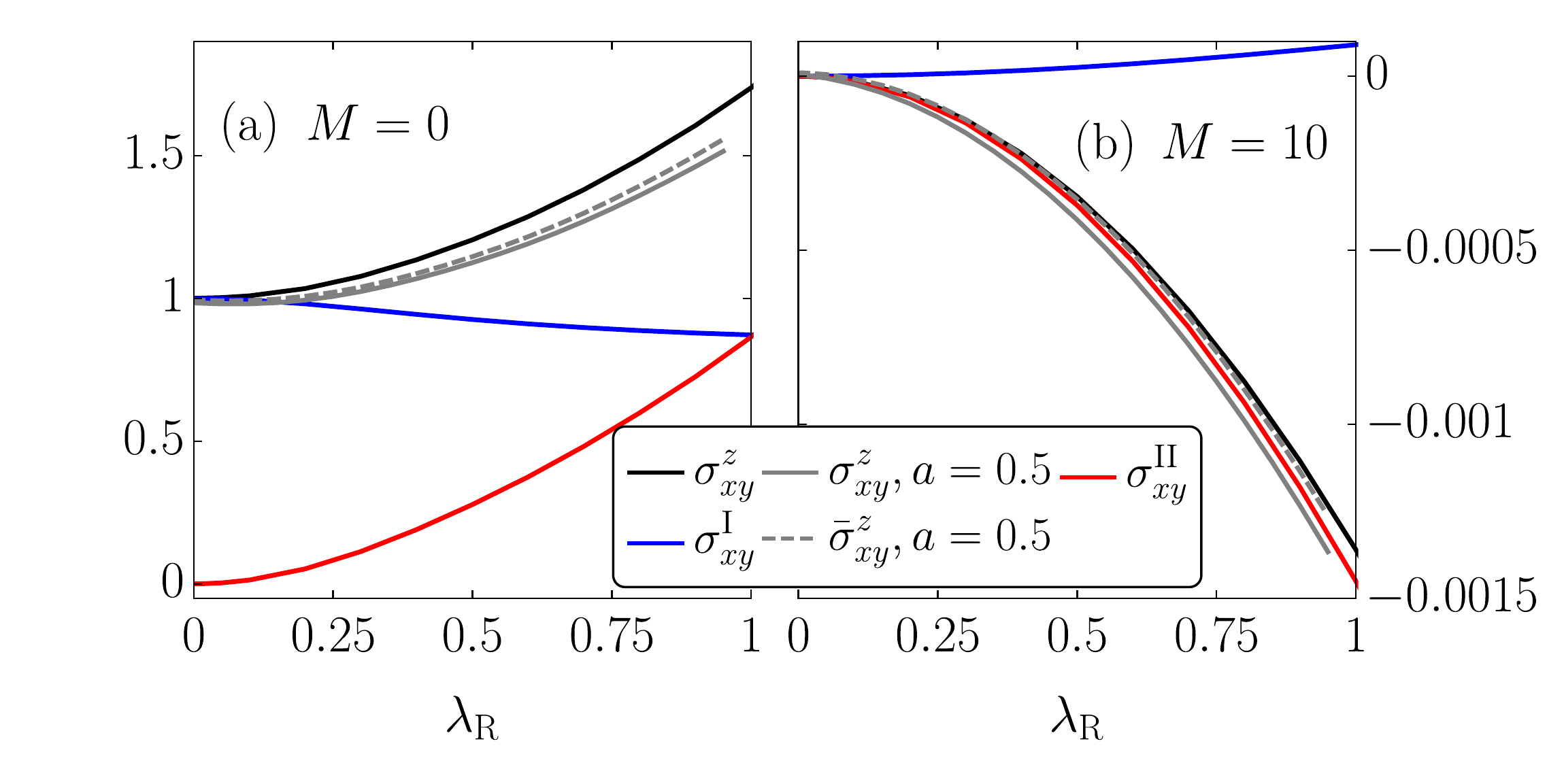}\includegraphics[width=130pt]{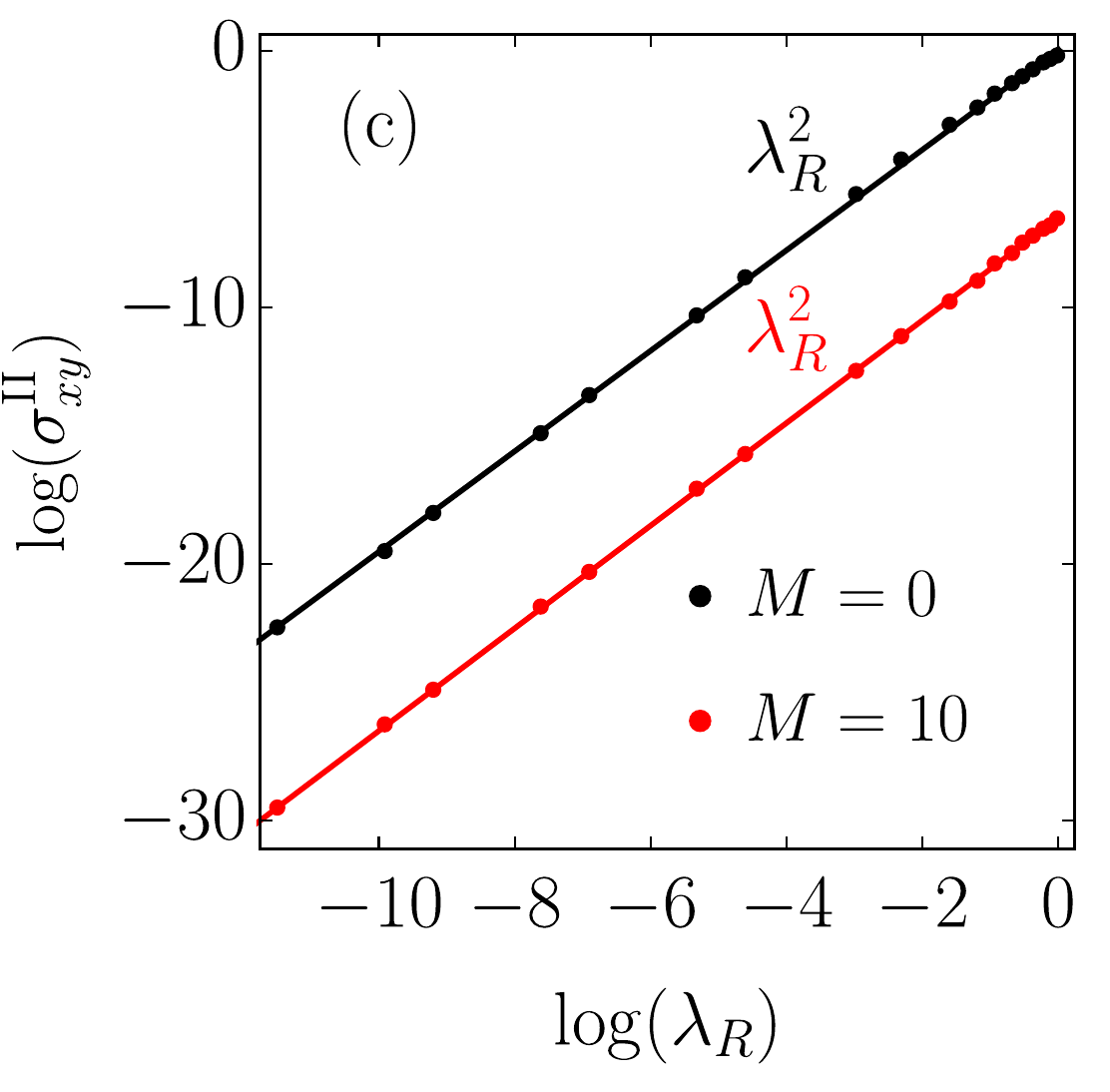}}
	\caption{Spin Hall conductivity $\sigma_{xy}^{z}$, the ``Chern-like'' contribution $\sigma_{xy}\su{I}$ and the ``extra'' contribution $\sigma_{xy}\su{II}$ in the KM model at $t = \lambda\sub{SO}=1$ on a $101 \times 101$ finite lattice. 
		Dependence on $\lambda\sub{R}$ for the system in (a) the topological phase at $M=0$ and (b) in the trivial phase at $M=10$. The grey plots correspond to the spin Hall conductivity, calculated by using the spin current operator $\mathbf{J}^z$ (solid) and the conventional spin current operator $\frac{1}{2}\{\dot{\mathbf{X}},S^z\}$ ($\bar{\sigma}_{xy}^z$, dashed), of a system with mirror-symmetry breaking terms of the amplitude $a=0.5$. (c) Scaling of $\sigma_{xy}\su{II}$ in $\lambda\sub{R}$ on a log-log scale. Dots show numerical results while full lines denote fitted $\lambda\sub{R}^2$ curve.}
	\label{fig:SxyOfR}
\end{figure*}

To compute the spin Hall conductivity as in Eq.~\eqref{eqn:I&IIxkKM}, we need to set $c_A(k_1)=1$, $c_B(k_1)=\eu^{\iu 2\pi/3 k_1}$ in Eq.~\eqref{eqn:partialFourier} and $x_A(R_1)=\frac{\sqrt{3}}{2}R_1$, $x_B(R_1)=\frac{\sqrt{3}}{2}R_1+\frac{1}{2\sqrt{3}}$ in Eq.~\eqref{eqn:Xsigma}.  The numerical results are illustrated in Fig.~\ref{fig:SxyOfR}. For $\lambda\sub{R}=0$, the spin is conserved and the spin Hall conductivity is exactly $0$ in the trivial phase at $M=10$ and $1$ in the topological phase at $M=0$. As in the BHZ model, $\sigma_{xy}\su{II}$ grows quadratically with $\lambda\sub{R}$, while $\sigma_{xy}\su{I}$ stays approximately unchanged. We compare these results with the ones given by the conventional definition of the spin current, $\bJ^z\sub{conv} = \frac{1}{2} \{ \dot{\bX}, S^z\}$, and observe that the two different definitions give the same value of the spin Hall conductivity. As shown in Appendix~\ref{sec:conv=prop}, this equivalence is a manifestation of the mirror symmetry $(x,y) \mapsto (-x,y)$,
\begin{equation} \label{hidden}
H_0(k_x,k_y)^* = H_0(-k_x,k_y),
\end{equation} 
which transforms the lattice into a new hexagonal lattice, shifted by $\bsd_2$ with respect to the original one. In order to test this hypothesis, we add to the Hamiltonian terms that break the mirror symmetry, while still preserving the time-reversal symmetry, namely
\begin{equation}
\begin{split}
&a\sin\left(\frac{\sqrt{3}k_x}{2}\right)\,\cos\left(\frac{k_y}{2}\right)\,\big[s^x\otimes\sigma^z+(s^x-s^y)\otimes\sigma^0\big]\\
&+a\cos\left(\frac{\sqrt{3}k_x}{2}\right)\,\sin\left(\frac{k_y}{2}\right)\,s^y\otimes\sigma^z.
\end{split}
\end{equation}
 The corresponding results, shown in Fig.~\ref{fig:SxyOfR}, clearly show that in the case of broken mirror symmetry the two definitions of the spin current yield different results, the difference becoming larger for larger values of $\lambda\sub{R}$. A different argument for this equivalence, relying instead on the hexagonal symmetry of the KM model, is presented in \cite{MarcelliMonacoPanatiTeufel}.

	\section{Conclusions}
	
In this paper, we presented a formula for the spin Hall conductivity in 2D band insulators, following the strategy employed in Ref.~\cite{MarcelliMonacoPanatiTeufel} (which, even though we chose to work with lattice Hamiltonians, applies also to systems in the continuum). 
The spin current operator is modelled according to Ref.~\cite{Niu06}; it satisfies a continuity equation and the Onsager relations even when the spin is not conserved. The spin Hall conductivity was shown to consist of a ``Chern-like'' contribution, which was also studied in Ref.~\cite{MarcelliPanatiTauber19}, and an extra term, which is non-zero in the presence of spin non-conserving terms in the Hamiltonian. This splitting of the spin conductivity into two contributions differs slightly from the one performed in Ref.~\cite{MarcelliMonacoPanatiTeufel}, where instead the two terms arise from writing the spin current operator $\bJ^z$ as the sum of the operator $\iu [H_0,\bX] S^z$, which appears also in the conventional definition of the spin current, plus the rest, as in Eq.~\eqref{eq:Scurrent}. Our ``Chern-like'' term, instead, identifies the perturbing potential $-E \, Y$ and the current operator $J_x^z = \iu [H_0, Q_x]$, respectively by the appearance of the operators $[Y,\Pi_0]$ and $[Q_x, \Pi_0]$ in the definition \eqref{eqn:I&II} of $\sigma_{xy}^{\mathrm{I}}$. Moreover, $\sigma_{xy}^{\mathrm{I}}$ was also investigated in Ref.~\cite{MarcelliPanatiTauber19} in relation with the spin Hall conductance.

We illustrated  the formula for the spin Hall conductivity by implementing it numerically in the BHZ model and the KM model for time-reversal symmetric topological insulators. Interestingly, the ``Chern-like'' contribution to the spin Hall conductivity stays close to the quantized value for both models. At any rate, the total spin Hall conductivity deviates from the quantized value as soon as the conservation of spin is broken.

The same results for the BHZ model were obtained in Ref.~\cite{Ulcakar18}, where the electric field is turned on through an adiabatic time modulation and the spin response is calculated via the conventional spin current operator $\frac{1}{2} \{ \dot{\bX}, S^z\}$. We showed that the resulting conductivities agree due to additional spatial symmetries that are present in the analysed models.

Complementary to the presentation of this paper, expressions for the spin Hall conductivity were also derived in the form of a St\v{r}eda formula in Ref.~\cite{YangChang06} starting from the conventional spin current operator, and in Ref.~\cite{Murakami06} from the one that satisfies the continuity equation. In the insulating regime, Ref.~\cite{YangChang06} identifies the contributions in the conductivity associated to spin-conservation breaking terms in the Hamiltonian (the same role played in our formulae by $\sigma^{\mathrm{II}}_{xy}$, see Eq.~\eqref{eqn:I&II}); in contrast, the use of the spin current operator $\iu [H_0, X S^z]$ allows the author of Ref.~\cite{Murakami06} to express the spin Hall conductivity in a more compact form, and to claim that its value is almost quantized in the KM model also in presence of Rashba  interactions. Our findings resolved the deviation of the spin Hall conductivity from the quantized value, which in the KM scales quadratically in the strength of the Rashba spin-orbit coupling.

\begin{acknowledgments}
We thank G.~Marcelli, G.~Panati, R.~Raimondi and S.~Teufel for useful discussions and their valuable comments on an early version of this paper. This work has been supported by the European Research Council (ERC) under the European Union’s Horizon 2020 research and innovation programme (ERC CoG UniCoSM, grant agreement n.724939). L. Ul\v{c}akar acknowledges the support by the Slovenian Research Agency under contract no. P1-0044 and L'Or\'eal-UNESCO For Women in Science Programme.
\end{acknowledgments}

\appendix
	
\section{Trace per unit volume} \label{app:TPUV}

The finite Bravais lattice $\Gamma_{N_1, N_2}$ is generated by the basis $\set{\bsa_1, \bsa_2}$ as $\br = R_1 \bsa_1 + R_2 \bsa_2$, with $R_j \in \set{-(N_j-1)/2, \ldots, (N_j-1)/2}$ for $j \in \set{1,2}$. Both translation-invariant operators and operators of the form $\bX B$ with translation-invariant $B$ have a well-defined TPUV
\begin{equation}  \label{eqn:tau_def}
\begin{aligned}
\tau\{A\} & = \lim_{N_1, N_2 \to \infty} \frac{1}{|\Omega| \, N_1 \, N_2} %\\ 
%& \quad 
\sum_{\br \in \Gamma_{N_1,N_2}} %\sum_{\bx \in \Omega} 
\sum_{s, \sigma} \bra{%\bx-
\br \, s \, \sigma} A \ket{%\bx-
\br \, s \, \sigma}%\nonumber,
.
\end{aligned}
\end{equation}
%where the sum over $\bx$ extends over the fundamental cell $\Omega$ of the crystal. 
Moreover, due to the fact that any finite lattice with an odd number of cells in each direction is symmetric under inversion around the origin, the quantity in the above limit is exactly independent of $N_1$ and $N_2$, and the TPUV of any such operator $A$ can be computed as%~\cite{MarcelliMonacoPanatiTeufel}
\begin{equation} \label{eqn:TPUV} 
\tau\{A\} = \frac{1}{|\Omega|} %\sum_{\bx \in \Omega} 
\sum_{s, \sigma} \bra{\mathbf{0} \, s \, \sigma} A \ket{\mathbf{0} \, s \, \sigma}.
\end{equation}

This observation is clear for translation-invariant $A$, so we prove it for $A = \bX B$, following an argument presented in \cite{MarcelliPanatiTauber19, MarcelliMonacoPanatiTeufel}. To this end, first observe the following commutation relation:
\begin{equation} \label{eqn:[X,T]}
\big[\bX , T_{\br}\big] = \br \, T_{\br},
\end{equation}
where $T_{\br} \ket{\bx' \, s \, \sigma} = \ket{\bx'+\br \, s \, \sigma}$ is the translation operator with respect to the Bravais lattice vector $\br$. The summands that define the trace per unit volume in Eq.~\eqref{eqn:tau_def} of $A = \bX B$ reduce then to
\begin{align}
&\bra{\mathbf{0} \, s \, \sigma} T_{\br}^\dagger \, \bX B \, T_{\br} \ket{\mathbf{0} \, s \, \sigma} = \bra{\mathbf{0} \, s \, \sigma} T_{\br}^\dagger \, \bX \, T_{\br} \, B \ket{\mathbf{0} \, s \, \sigma}\nonumber \\
& = \bra{\mathbf{0} \, s \, \sigma} \bX B \ket{\mathbf{0} \, s \, \sigma} + \br \bra{\mathbf{0} \, s \, \sigma} B \ket{\mathbf{0} \, s \, \sigma}.
\end{align}
The first equality is due to the translation invariance of $B$, while the second is due to the commutation relation in Eq.~\eqref{eqn:[X,T]} and unitarity $T_{\br}^\dagger T_{\br} = \mathbf{1}$. When the second term on the right-hand side of the above equality is summed over $\br \in \Gamma_{N_1,N_2}$, the sum vanishes, as for each lattice vector $\br$ also $-\br$ is in the finite lattice.

For a translation-invariant operator $A$, which admits a Fourier representation $A(\bk)$, one has
\begin{equation}  \label{diagonal element}
%\begin{aligned}
\bra{\mathbf{0} \, s \, \sigma} A \ket{\mathbf{0} \, s \, \sigma} %& = \frac{1}{N_1 N_2} \sum_{\bk, \bk' \in \BZ} \eu^{-\iu(\bk-\bk')\cdot\bx} \bra{\bk' \, s \, \sigma} A \ket{\bk \, s \, \sigma} \\
%& = \frac{1}{N_1 N_2} \sum_{\bk, \bk' \in \BZ} \eu^{-\iu(\bk-\bk')\cdot\bx} \, \delta(\bk-\bk') \, A(\bk)_{s \, \sigma}^{s \, \sigma} \\
%& 
= \frac{1}{N_1 N_2} \sum_{\bk \in \BZ} A(\bk)_{s \, \sigma}^{s \, \sigma}%.
,
%\end{aligned}
\end{equation}
%The right-hand side is independent of $\bx$, 
and therefore the TPUV in Eq.~\eqref{eqn:TPUV} can be also computed as
\begin{equation}
%\begin{split}
\tau\{A\} %& 
= \frac{1}{|\Omega| \, N_1 \, N_2} \sum_{\bk \in \BZ} \sum_{s,\sigma} A(\bk)_{s \, \sigma}^{s \, \sigma} %\\
%& = \frac{1}{|\Omega| \, N_1 \, N_2} \sum_{\bk \in \BZ} \Tr_{\C^n}(A(\bk))
%\end{split}
\end{equation}
which in the thermodynamic limit $N_1, N_2 \to \infty$ reduces~to
\begin{equation} 
\tau\{A\} = \frac{1}{(2\pi)^2} \int_{\BZ} \di \bk \, \Tr_{\C^n}(A(\bk)). 
\end{equation}
From these expressions it can be inferred at once that $\tau\{A B\} = \tau\{B A\}$ for translation-invariant operators $A, B$. This cyclicity property is in general broken if one applies it instead to non-translation-invariant operators of the form $\bX B$ of the type considered in the main text.

\section{Linear response for spin currents} \label{app:internal}

%\subsection{For the ``proper'' spin current operator}

In order to calculate the spin current induced by the electric field, see Eq.~\eqref{eq:sHall}, we rewrite $ \bJ^z \, \Pi_1$ by using the Leibnitz rule for commutators $[A,BC] = [A,B]C + B[A,C]$ and the fact that $A\su{OD} = \big[[A,\Pi_0], \Pi_0 \big]$, as
\begin{equation} \label{B1}
\begin{aligned}
\iu [H_0,\bQ] \Pi_1 & = \iu \left[H_0,\bQ\su{D}\right] \Pi_1 + \iu \left[H_0,\bQ\su{OD}\right] \Pi_1  \\
& = \bD_1 + \iu \left[H_0, \bQ\su{OD} \Pi_1\right] - \iu \bQ\su{OD} [\hat{\bE} \cdot \bX,\Pi_0] \\
& = \bD_1 + \bD_2 -\iu \big[ [\bQ,\Pi_0], \Pi_0 \big] [\hat{\bE} \cdot \bX,\Pi_0]\\
& = \bD_1 + \bD_2 - \iu \big[ [\bQ,\Pi_0], \Pi_0 [\hat{\bE} \cdot \bX,\Pi_0] \big] \\
& \quad + \iu \Pi_0 \big[ [\bQ,\Pi_0],[\hat{\bE} \cdot \bX,\Pi_0] \big]\\
& = \bD_1 + \bD_2 + \bD_3 +\bK.
\end{aligned}
\end{equation}
In the above, we have set
\begin{equation}
\begin{aligned}
\bD_1 & = \iu \left[H_0,\bQ\su{D}\right] \Pi_1, \\
\bD_2 & = \iu \left[H_0, \bQ\su{OD} \Pi_1\right], \\
\bD_3 & = - \iu \big[ [\bQ,\Pi_0], \Pi_0 [\hat{\bE} \cdot \bX,\Pi_0] \big], \\
\bK & = \iu \Pi_0 \big[ [\bQ,\Pi_0],[\hat{\bE} \cdot \bX,\Pi_0] \big].
\end{aligned}
\end{equation}
Finally we set $\bD = \bD_1 + \bD_2 + \bD_3$, so that the definitions of $\bK$ and $\bD$ coincide with the ones in Eq.~\eqref{K&D}. Notice that $[A,B]^\dagger = - [A^\dagger, B^\dagger]$ (which implies $(\iu [A,B])^\dagger = \iu [A^\dagger, B^\dagger]$), and hence $\bK$ is self-adjoint:
\begin{equation}
\begin{split}
\bK^\dagger & = \iu \big[ [\bQ,\Pi_0]^\dagger,[\hat{\bE} \cdot \bX,\Pi_0]^\dagger \big] \Pi_0 \\
& = \iu \big[ -[\bQ,\Pi_0],-[\hat{\bE} \cdot \bX,\Pi_0] \big] \Pi_0 \\
& = \iu \Pi_0 \big[ [\bQ,\Pi_0],[\hat{\bE} \cdot \bX,\Pi_0] \big] = \bK,
\end{split}
\end{equation}
where in the second-to-last equality we used that operators of the form $[A, \Pi_0]$ are off-diagonal, and therefore the commutator of $[\bQ,\Pi_0]$ and $[\hat{\bE} \cdot \bX,\Pi_0]$ is diagonal, thus commuting with $\Pi_0$. Therefore in particular $\tau\{\bK\}$ is real-valued.

\section{Agreement of ``conventional'' and ``proper'' spin Hall conductivities} \label{sec:conv=prop}

%\textbf{[We have to understand what we want to do with these considerations]}

The conventionally used spin current operator is defined as 
\begin{equation}
 \bJ^z\sub{conv} = \frac{1}{2} \big\{ \dot{\bX}, S^z \big\}  = \frac{1}{2} \big\{ \iu [H_0,{\bX}], S^z \big\}.
\end{equation}
%Assume that $\bJ^z = \bJ^z\sub{prop}$. 
If we split the %``proper'' 
spin current operator adopted from Ref.~\cite{Niu06} according to
\begin{equation} \label{eqn:convVSprop}
\bJ^z = \bJ^z\sub{conv} + \frac{1}{2} \, \big\{ \bX , \dot{S}^z \big\},
\end{equation}
and rewrite Eq.~\eqref{eqn:Jzrho} as
\begin{equation} \label{eqn:Jzrho1}
\langle \bJ^z \rangle_{\Pi}= \frac{1}{2 |\Omega|} %\sum_{\bx \in \Omega}
 \sum_{s, \sigma} \bra{\mathbf{0} \, s \, \sigma} \big\{ \bJ^z, \Pi \big\} \ket{\mathbf{0} \, s \, \sigma},
\end{equation}
we obtain
\begin{equation} \label{difference}
\begin{aligned}
&\re \: \tau \{ \bJ^z \, \Pi_1 \} = \re \: \tau \{ \bJ^z\sub{conv} \, \Pi_1 \} \\
& \quad + \frac{1}{4 |\Omega|} %\sum_{\bx \in \Omega} 
\sum_{s, \sigma} \bra{\mathbf{0} \, s \, \sigma} \left\{ \big\{ \bX, \dot{S}^z \big\}, \Pi_1 \right\} \ket{\mathbf{0} \, s \, \sigma}.
\end{aligned}
\end{equation}
We notice the following operator identity:
\begin{equation}
\left\{ \big\{ \bX, \dot{S}^z \big\} , \Pi_1 \right\} = \left\{ \big\{ \dot{S}^z, \Pi_1 \big\} , \bX \right\} + \left[ \dot{S}^z , \big[ \bX, \Pi_1 \big] \right] .
\end{equation}
The commutator on the right-hand side is a translation-invariant operator, and hence does not contribute to the TPUV because of cyclicity, $\tau\{AB\} = \tau\{BA\}$. Instead, in the expectation of the summand $\big\{ \dot{S}^z, \Pi_1 \big\} \, \bX + \bX \, \big\{ \dot{S}^z, \Pi_1 \big\}$, the position operator $\bX$ will act on the state $\ket{\mathbf{0} \, s \, \sigma}$ (or the corresponding $\bra{\mathbf{0} \, s \, \sigma}$). If this state is localized at $\bx = \mathbf{0}$, also this expectation will vanish: this is what happens in the BHZ model. 

In general, like in the hexagonal KM model, there will be other sites in the unit cell contributing to the above TPUV. In this case, the position operator $\bX$ acts as $\bX \ket{\mathbf{0} \, s \, \sigma} = \mathbf{r}_\sigma \, \ket{\mathbf{0} \, s \, \sigma} $, where $\mathbf{r}_\sigma$ is a displacement vector ($\mathbf{r}_A = \mathbf{0}$ and $\mathbf{r}_B = \bsd_2$ in the KM model), and
\begin{equation}
\begin{split}
&\bra{\mathbf{0} \, s \, \sigma} \left\{ \big\{ \dot{S}^z, \Pi_1 \big\} , \bX \right\} \ket{\mathbf{0} \, s \, \sigma} \\
&=2 \, \mathbf{r}_\sigma \bra{\mathbf{0} \, s \, \sigma} \big\{ \dot{S}^z, \Pi_1 \big\} \ket{\mathbf{0} \, s \, \sigma}.
\end{split}
\end{equation}
Using the Leibnitz rule for the commutator, the operator appearing on the right-hand side can be rewritten as
\begin{equation} \label{eqn:LeibnitzRule}
\begin{aligned}
\big\{ \iu [H_0,S^z], \Pi_1 \big\} & = \iu \big[ H_0, \{  S^z, \, \Pi_1\} \big] - \big\{ S^z, \, \iu [H_0, \Pi_1] \big\} \\
& = \iu \big[ H_0, \{  S^z, \, \Pi_1\} \big] - \big\{ S^z, \, \iu [\hat{\bE} \cdot \bX, \Pi_0] \big\} \\
& = \iu \big[ H_0, \{  S^z, \, \Pi_1\} \big] - \iu \, \hat{\bE} \cdot \big[ \bX, \{ S^z, \Pi_0 \} \big],
\end{aligned}
\end{equation}
using Eq.~\eqref{eqn:CommPi1} in the second equality and the fact that the position operator and the spin operator commute in the third equality. On the right-hand side of the above, the second summand does not have diagonal elements in the $\ket{\bx \, s \, \sigma}$ basis: in particular
\begin{align}
& \bra{\mathbf{0} \, s \, \sigma} \big[\bX, \{ S^z, \Pi_0 \} \big] \ket{\mathbf{0} \, s \, \sigma} = \mathbf{r}_\sigma \bra{\mathbf{0} \, s \, \sigma} \{ S^z, \Pi_0 \} \ket{\mathbf{0} \, s \, \sigma}\nonumber\\
&  -  \bra{\mathbf{0} \, s \, \sigma} \{ S^z, \Pi_0 \} \ket{\mathbf{0} \, s \, \sigma} \mathbf{r}_\sigma = 0.
\end{align}

Call $B_1 = \iu \big[ H_0, \{  S^z, \, \Pi_1\} \big]$: it is a translation invariant operator, and therefore (compare Eq.~\eqref{diagonal element})
\begin{equation}
\bra{\mathbf{0} \, s \, \sigma} B_1 \ket{\mathbf{0} \, s \, \sigma} %& = \frac{1}{N_1 \, N_2} \sum_{\bk, \bk' \in \BZ} \eu^{-\iu(\bk-\bk')\cdot\mathbf{0}} \bra{\bk' \, s \, \sigma} B_1 \ket{\bk \, s \, \sigma} %\\
%& = \frac{1}{N_1 \, N_2} \sum_{\bk, \bk' \in \BZ} \eu^{-\iu(\bk-\bk')\cdot\bx} \, \delta(\bk-\bk') \, B_1(\bk)_{s \, \sigma}^{s \, \sigma} \\
%&
= \frac{1}{|\Omega| \, N_1 \, N_2} \sum_{\bk \in \BZ} B_1(\bk)_{s \, \sigma}^{s \, \sigma}.
\end{equation}
The difference $\re \: \tau \{ \bJ^z \, \Pi_1 \} - \re \: \tau \{ \bJ^z\sub{conv} \, \Pi_1 \}$ then equals
\begin{equation} \label{B_1}
\frac{1}{2} \, \frac{1}{|\Omega| \, N_1 \, N_2} \sum_{\sigma} \mathbf{r}_\sigma \sum_{\bk \in \BZ} \sum_{s} B_1(\bk)_{s \, \sigma}^{s \, \sigma}.
\end{equation}
%which can be computed directly in the thermodynamic limit $N_1, N_2 \to \infty$ as
%\begin{equation} \label{eqn:Prop-Conv}
%\frac{1}{2} \, \int_{\BZ} \frac{\di \bk}{(2 \pi)^2} \left( \sum_{s, \sigma} \mathbf{r}_\sigma \, B(\bk)_{s\sigma}^{s\sigma} \right).
%\end{equation}
%where $\Tr_{\C^2}(\cdot)$ is the trace over the spin degrees of freedom~$s$. 

%We now look at the structure of the operator $B(\bk) = \{ \iu [H_0(\bk), S^z], \, \Pi_1(\bk)\}$. Using the Leibnitz rule for the commutator, it can be rewritten as
%\begin{align*}
%B(\bk) & = \iu \big[ H_0(\bk), \{  S^z, \, \Pi_1(\bk)\} \big] - \big\{ S^z, \, \iu [H_0(\bk), \Pi_1(\bk)] \big\} \\
%& = \iu \big[ H_0(\bk), \{  S^z, \, \Pi_1(\bk)\} \big] + \big\{ S^z, \, \hat{\bE} \cdot \nabla_{\bk} \Pi_0(\bk) \big\}
%\end{align*}
%using the $\bk$-space version of~\eqref{eqn:CommPi1} in the second equality. The second term on the right-hand side of the above contributes to~\eqref{eqn:Prop-Conv} for
%\[ \frac{1}{2} \, \int_{\BZ} \frac{\di \bk}{(2 \pi)^2} \hat{\bE} \cdot \nabla_{\bk} \left( \sum_{s, \sigma} \mathbf{r}_\sigma \, \big[ \{ S^z, \, \Pi_0(\bk) \} \big]_{s \, \sigma}^{s \, \sigma} \right) \]
%which vanishes because the integrand is a derivative of a periodic function of $\bk$, and the integration is performed on the Brillouin torus. 

%\textbf{[I don't know how to treat this term!]}

We now exploit the mirror symmetry, shown in Eq.~\eqref{hidden}, of the KM model. Notice first of all that it is inherited by the Fermi projection:
\begin{equation}
\Pi_0(k_x,k_y)^* = \Pi_0(-k_x,k_y).
\end{equation}
With this one can argue that, since the solution to the Eqs.~\eqref{eqn:Pi1OD} and~\eqref{eqn:CommPi1} is unique, then also $\Pi_1(\bk)$ satisfies the same relation. Consequently, as $S^z$ has real components,
\begin{equation}
\begin{split}
B_1(k_x,k_y)^* & = \big( \iu \big[ H_0(k_x,k_y), \{  S^z, \, \Pi_1(k_x,k_y) \} \big] \big)^* \\
& = - B_1(-k_x,k_y),
\end{split}
\end{equation}
so that the expression in Eq.~\eqref{B_1} is odd in $k_x$, and thus sums to zero over the BZ.

We conclude finally that
\begin{equation}
\re \tau \{ \bJ^z \, \Pi_1 \} = \re \tau \{ \bJ^z\sub{conv} \, \Pi_1 \} ,
\end{equation}
and that the spin conductivity tensor $\boldsymbol{\sigma}^z$ is independent of the choice of spin current operator both in the BHZ and in the KM model.

\end{document}